\documentclass[prl,aps,floatfix,amsmath,amssymb,superscriptaddress,tightenlines,twocolumn]{revtex4-1}

\usepackage{graphicx}
\usepackage{epstopdf}
\newcommand{\be}{\begin{equation}}
\newcommand{\ee}{\end{equation}}
\usepackage{amsmath}
\usepackage{amsfonts}
\usepackage{bm}
\usepackage{color}
\usepackage{subfigure}
\usepackage{amsthm}
\usepackage[latin1]{inputenc}
\usepackage{tikz}
\usepackage{comment}

\newcommand{\ket}[1]{\left|#1\right\rangle}

\newcommand{\Tr}{\textrm{Tr}}

\begin{document}
\title{Holographic quantum simulation}
\author{Isaac H. Kim}
\affiliation{IBM T.J. Watson Research Center}
\date{\today}

\begin{abstract}
	A one-dimensional quantum simulator can simulate two-dimensional quantum many-body systems. A representation of a low-energy state is obtained by applying a feedback loop.
\end{abstract}
\maketitle

{\it Introduction---}
Most modern-day computers are universal\cite{Turing1936}, but this was not always the case. In the early days of computing, machines were built to solve a rather specific family of problems. A paradigmatic example is the differential analyser, which was capable of solving certain low-order differential equations\cite{Bush1927}. Despite being an analog device, it boasted an average error rate of one  percent, which was often accurate enough for certain scientific purposes, e.g., simulation of transmission lines\cite{BUSH1931}. This is a telling anecdote; a noisy, but a sufficiently accurate device could resolve an intractable scientific problem at the time.

The same can be said about quantum computers. While quantum algorithms\cite{Shor1997,Grover1996,Lloyd1996}  typically require a fault-tolerant quantum computer, certain problems may be amenable to a noisy device. This is especially so for simulating quantum many-body systems that appear in nature, a vision that was famously put forward by Feynman\cite{Feynman1982}. Recent progress in quantum technology provides an ample amount of opportunities to develop and test these ideas in near-term experiments\cite{Kim2010,Simon2011,Trotzky2012,Greif2013,Hart2015}. 

An important milestone would be the simulation of strongly correlated two-dimensional(2D) systems, e.g., the Hubbard model\cite{Hubbard1963,Anderson1987}. Indeed, mapping the phase diagram of such a model remains challenging; development of new numerical methods is an active area of ongoing research\cite{White1992,Verstraete2004a,Vidal2008}; analog simulation based on ultracold atoms are facing a challenge in cooling\cite{McKay2011}; and the architectures for digital quantum simulators\cite{Kim2010,Lanyon2011,Blatt2012,Islam2012,Houck2012,Barends2015a,Barends2015,Smith2015} remain one-dimensional(1D) or small-scale.

The purpose of this paper is to propose a method that can overcome these limitations of digital simulators. We argue that the existing 1D architecture is sufficient to simulate certain 2D systems. We propose an experimental protocol, and provide a theoretical justification.

If we were to simulate arbitrary states in the Hilbert space this way, we would inevitably fail. However, we expect the physically relevant states to occupy only a small fraction of the Hilbert space\cite{Hastings2007,Eisert2010}. In order to understand the boundary between the physical and the unphysical states, we derive a universal structure of ground state entanglement that is present in systems with a mass gap. Such systems are expected to obey the area law of entanglement entropy. While the area law in the sense of having an area-like upper bound is unlikely to lead to an efficient representation\cite{Ge2014}, they do if we demand a more refined structure\cite{Kitaev2006,Levin2006,Kim2014}. Any states equipped with such a structure can be represented by a history of a 1D quantum simulator.

Owing to this universal structure, the 1D quantum simulator can serve as a ``variational machine'' that can probe the physics of 2D quantum many-body systems with a mass gap\footnote{In systems with Goldstone modes, we expect the noise to effectively act as a symmetry-breaking term. The noise rate is likely to determine the mass.}. A low-energy configuration can be found by estimating the energy of the simulated system and adjusting the simulator to lower the energy. Once a minimum-energy configuration is found, the relevant order parameters can 
be studied.

That the state of the device can be used as an ansatz was put forward in Ref.\cite{Peruzzo2013}. Our work shows that the dimension of the ansatz need not be limited by the size of the device. By studying the history of the device, as opposed to a static state, one can study a much larger system. Indeed, any device capable of implementing a universal gate set in a quasi-1D architecture can simulate gapped 2D quantum many-body system, similar to the proposal in Ref.\cite{Osborne2010,Barrett2013}.

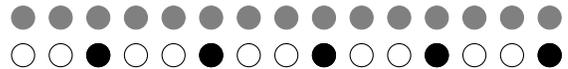
\begin{figure}[h] 
	\begin{tikzpicture}
		\foreach \i in {1,...,15}
		{
			\node[circle,draw=gray,fill=gray](\i) at (0.5 * \i, 0.5) {};
		}
		\foreach \i in {1,...,15}
		{
			\node[circle,draw=black,fill=white](\i) at (0.5 * \i, 0.0) {};
		}
		\foreach \i in {1,...,5}
		{
			\node[circle,draw=black,fill=black](\i) at (1.5 * \i, 0.0) {};
		}
	\end{tikzpicture}
	\caption{ The qubits are arranged so that the system(black) qubits are surrounded by bath(white) and sink(gray) qubits. Each of the qubits can	interact with their neighbors. \label{fig:architecture}}
\end{figure}

\textit{Architecture---}
Our proposal can be implemented on platforms equipped with quasi-1D array of qubits with tunable nearest-neighbor interactions; see FIG.\ref{fig:architecture} for one such arrangement\footnote{One can make the architecture to be truly 1D with the expense of allowing slightly nonlocal interaction, e.g., next-nearest-neighbor interaction.}. The interactions should be rich enough that a universal 
gate set can be implemented.

The qubits are classified into three categories: system, bath, and sink. Each of these qubits play a different role, and they can be made differently to suit these purposes. Experimentalists will extract the information about the simulated system from the system qubits. These qubits must be equipped with accurate measurements. The bath qubits carry the quantum correlation between system qubits at different times. Decoherence must be minimized; otherwise, the simulator will be unable to properly simulate this quantum correlation. On the other hand, measurement need not be perfect because it is never directly measured throughout the experiment. The sink qubits are constantly reset to a fixed initial state. As such, it requires a fast initialization routine.

\textit{State of a 2D system---}
How can a 1D simulator simulate a 2D system? In order to answer this question, we need to ask a somewhat philosophical question: what is a quantum state? Mathematically, it is a positive semi-definite functional with a unit trace. A state is defined by (i) specifying the expectation values for all the observables and (ii) ensuring that the functional is positive semi-definite and normalized. Thus, any entity that can supply the information about the expectation values effectively has a description of a quantum state, provided that the entity can convincingly prove the second condition.

This abstract entity, in this paper, would be the 1D quantum simulator. The expectation values are expressed in terms of a data that can be gathered from the 1D simulator. The functional will be positive semi-definite and normalized by the nature of the 1D simulator, even in the presence
of noise.

Any expectation value can be expressed in terms of a linear combination of $n$-point functions. Therefore, every expectation value is determined by specifying every $n$-point function.
The $n$-point functions of the simulated system are defined in terms of the \emph{time-dependent} $n$-point functions of the simulator:
\begin{equation}
	\langle O_{(x_1,y_1)}\cdots O_{(x_n,y_n)} \rangle_{\text{2D}} := \langle O_{x_1}(y_1)  \cdots O_{x_n}(y_n) \rangle_{\text{1D}}, \label{eq:state_definition},
\end{equation}
where $O_{(x,y)}$ is an operator acting on the qubit at location $(x,y)$ of the 2D system and $O_x$ is an operator acting on the \emph{system qubit} at location $x$ of the 1D simulator. Also,
$\langle \cdots \rangle_{\text{2D}}$ is the expectation value of the observables for the 2D system and $\langle \cdots \rangle_{\text{1D}}$ can be measured by destructively measuring the
observables $O_{x_i}$ at time $y_i$. For example, to measure $\langle \sigma^x_{(x_1,y_1)} \sigma^z_{(x_2,y_2)} \rangle_{\text{2D}}$, an experimentalist measures the system qubit at $x_1$
in the eigenbasis of $\sigma^x$ at time $y_1$, measure the system qubit at $x_2$ in the eigenbasis of $\sigma^z$ at time $y_2$, multiply the observed values, and repeat the measurements
until the statistical error becomes sufficiently small.

\begin{figure}[h]
	\begin{tikzpicture}[scale=1.2]
		\foreach \x in {0,0.3,...,1.5}{
			\foreach \y in {0,0.3,...,1.5}{
				\draw[black] plot [only marks, mark size=2.5, mark=*] coordinates {(\x,\y)};
		}
		}
		\draw[xshift=-0.15cm,yshift=-0.15cm] (0,0)--(1.5,0)--(1.5,0.3)--(0,0.3)--(0,0);
		\draw[thick,->] (1.5,0.6)--(1.75,0.6);
		\begin{scope}[xshift=2cm]
		\foreach \x in {0,0.3,...,1.5}{
			\foreach \y in {0,0.3,...,1.5}{
				\draw[black] plot [only marks, mark size=2.5, mark=*] coordinates {(\x,\y)};
		}
		}
		\draw[xshift=-0.15cm,yshift=-0.15cm] (0,0.3)--(1.5,0.3)--(1.5,0.6)--(0,0.6)--(0,0.3);
		\draw[thick,->] (1.5,0.6)--(1.75,0.6);
		\end{scope}
		\begin{scope}[xshift=4cm]
		\foreach \x in {0,0.3,...,1.5}{
			\foreach \y in {0,0.3,...,1.5}{
				\draw[black] plot [only marks, mark size=2.5, mark=*] coordinates {(\x,\y)};
		}
		}
		\draw[xshift=-0.15cm,yshift=-0.15cm] (0,0.6)--(1.5,0.6)--(1.5,0.9)--(0,0.9)--(0,0.6);
		\end{scope}

	\end{tikzpicture}
	\caption{At each time, the auxiliary 1D system(box) interacts with each rows of qubits that are enclosed within the box. The auxiliary system is the composite of bath and sink qubits. Each of the rows are system qubits at different times.\label{fig:gliding}}
\end{figure}
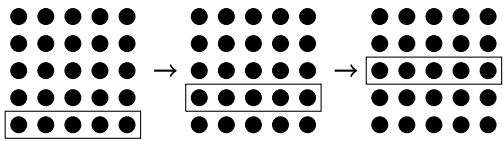
This prescription defines a valid quantum mechanical state, provided that all the system qubits are initialized to a fixed state after the measurement. To see this on an intuitive level, consider a set of qubits arranged on a 2D grid. These qubits are initialized to a fixed product state, and this state is transformed into another state by ``gliding'' an auxiliary 1D system over these qubits and letting them sequentially interact with each other; see FIG.\ref{fig:gliding}. A crucial property of this process is that the interaction between the auxiliary system and a row of qubits does not affect the measurement outcomes of the qubits that precede them. Therefore, instead of measuring an observable after completely transforming the state, one can measure each rows as soon as they interact with the auxiliary system. Going one step further, one can simply reset the row to the fixed state after performing the measurement, and let that state interact with the auxiliary system. The auxiliary system can be viewed as the bath-sink composite and each of the rows can be viewed as the system qubits at different times. Because every operation in this procedure is physically allowed, the state implicitly defined by Eq.\ref{eq:state_definition} is valid.

More formally, let us denote the system at time $t$ as $S_t$, $B$ as the bath, and $S'$ as the sink. The physical process of initializing the system and applying a circuit to the system-bath-sink composite can be represented by a quantum channel\cite{Nielsen2011}, $\Phi_t$, whose domain is $BS'$ and image is $S_tBS'$. Thus, we have a state that encapsulates the history of the system: 
\begin{equation}
	\rho = \Tr_{BS'}\circ \Phi_{\ell_y} \circ \cdots \Phi_0 [\rho^{BS'}].\label{eq:fcs}
\end{equation}
By the definition of channel, $\rho$ is a valid quantum mechanical state\cite{Fannes1992}.

\textit{Variational method---}
A 1D simulator can simulate 2D quantum many-body systems, by using Eq.\ref{eq:state_definition}. This prescription implies that a valid quantum mechanical state is defined in terms of the history of the simulator. Because the history is determined by the time evolution of the device, the state is defined by the quantum circuit applied to the simulator; the circuit elements parametrize the variational state.

The circuit can be arbitrary, but one rule should be enforced. At the beginning of each time steps, the system and the sink qubits need to be initialized to some fixed state, say $\ket{0}$. The initialization of the system qubit is to ensure that Eq.\ref{eq:state_definition} defines a valid quantum mechanical state. As for the sink qubits, our theoretical analysis seems to require it to be initialized at every time step. However, there is a logical possibility that this may be unnecessary. 

Provided that these rules are obeyed, a low-energy state can be found by a simple feedback loop. Energy is measured after applying a random perturbation to the circuits. The change is accepted only if the energy is lowered. Once the energy converges, relevant order parameters are measured.  

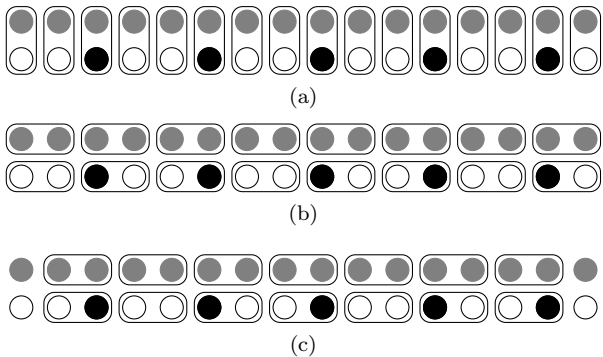
\begin{figure}[h]
	\subfigure[]{
	\begin{tikzpicture}
		\foreach \i in {1,...,16}
		{
			\draw[rounded corners] (0.5 * \i-0.2, -0.2) rectangle (0.5 * \i + 0.2, 0.7);
		}
		\foreach \i in {1,...,16}
		{
			\node[circle,draw=gray,fill=gray](\i) at (0.5 * \i, 0.5) {};
		}
		\foreach \i in {1,...,16}
		{
			\node[circle,draw=black,fill=white](\i) at (0.5 * \i, 0.0) {};
		}
		\foreach \i in {1,...,5}
		{
			\node[circle,draw=black,fill=black](\i) at (1.5 * \i, 0.0) {};
		}
	\end{tikzpicture}
}
	\subfigure[]{
	\begin{tikzpicture}
		\foreach \i in {1,...,8}
		{
			\draw[rounded corners] (1. * \i -0.7, 0.3) rectangle (1. * \i + 0.2, 0.7);
			\draw[rounded corners] (1. * \i -0.7, -0.2) rectangle (1. * \i + 0.2, 0.2);
		}
		\foreach \i in {1,...,16}
		{
			\node[circle,draw=gray,fill=gray](\i) at (0.5 * \i, 0.5) {};
		}
		\foreach \i in {1,...,16}
		{
			\node[circle,draw=black,fill=white](\i) at (0.5 * \i, 0.0) {};
		}
		\foreach \i in {1,...,5}
		{
			\node[circle,draw=black,fill=black](\i) at (1.5 * \i, 0.0) {};
		}
	\end{tikzpicture}
	}

	\subfigure[]{
	\begin{tikzpicture}
		\foreach \i in {1,...,7}
		{
			\draw[rounded corners] (1. * \i -0.7+0.5, 0.3) rectangle (1. * \i + 0.7, 0.7);
			\draw[rounded corners] (1. * \i -0.7+0.5, -0.2) rectangle (1. * \i + 0.7, 0.2);
		}
		\foreach \i in {1,...,16}
		{
			\node[circle,draw=gray,fill=gray](\i) at (0.5 * \i, 0.5) {};
		}
		\foreach \i in {1,...,16}
		{
			\node[circle,draw=black,fill=white](\i) at (0.5 * \i, 0.0) {};
		}
		\foreach \i in {1,...,5}
		{
			\node[circle,draw=black,fill=black](\i) at (1.5 * \i, 0.0) {};
		}
	\end{tikzpicture}
	}
	\caption{The circuit applies a sequence of two-qubit nearest-neighbor gates with fixed geometry. The sequence is chosen so that, by choosing an appropriate set of gates one can realize any quantum circuit of bounded depth.\label{fig:circuit}}
\end{figure}

\textit{Sufficiency of local gates---}
Circuits with long-range gates are difficult to implement in practice. Fortunately, for studying physical states, they can be ignored. Furthermore, within each time steps, the circuit depth is finite, independent of the system size; see FIG.\ref{fig:circuit}. 

\begin{figure}[h]
	\begin{tikzpicture}[scale=1.2]
		\draw (-0.5,0.3) node[] {$t$};
		\draw (-0.5,0.6) node[] {$t+1$};
		\draw (-0.5,0.9) node[] {$t+2$};
		\foreach \x in {0,0.3,...,1.5}{
			\draw[red] plot [only marks, mark size=2.5, mark=*] coordinates {(\x,0)};
			\foreach \y in {0.3,0.6,0.9}{
				\draw[black] plot [only marks, mark size=2.5, mark=*] coordinates {(\x,\y)};
			}
			\begin{scope}[xshift=-0.15cm, yshift=-0.15cm]
				\draw (\x+0.03,0)--(\x+0.3-0.03,0)--(\x+0.3-0.03,0.6)--(\x+0.03,0.6)--(\x+0.03,0);
			\end{scope}
		}
		\draw[thick,->] (1.5,0.5)--(1.75,0.5);
		\begin{scope}[xshift=2.7cm]
			\draw (-0.5,0) node[] {$t$};
			\draw (-0.5,0.6) node[] {$t+1$};
			\draw (-0.5,0.9) node[] {$t+2$};
			\foreach \x in {0,0.3,...,1.5}{
				\draw[red] plot [only marks, mark size=2.5, mark=*] coordinates {(\x,0.3)};
				\foreach \y in {0,0.6,0.9}{
					\draw[black] plot [only marks, mark size=2.5, mark=*] coordinates {(\x,\y)};
				}		
				\begin{scope}[xshift=-0.15cm, yshift=-0.15cm]
					\draw (\x+0.03,0.3)--(\x+0.3-0.03,0.3)--(\x+0.3-0.03,0.9)--(\x+0.03,0.9)--(\x+0.03,0.3);
				\end{scope}
		}
		\end{scope}
		\begin{scope}[xshift=2.7cm,yshift=-1.8cm]
			\draw (-0.5,0) node[] {$t$};
			\draw (-0.5,0.3) node[] {$t+1$};
			\draw (-0.5,0.9) node[] {$t+2$};
			\foreach \x in {0,0.3,...,1.5}{
				\draw[red] plot [only marks, mark size=2.5, mark=*] coordinates {(\x,0.6)};
				\foreach \y in {0,0.3,0.9}{
					\draw[black] plot [only marks, mark size=2.5, mark=*] coordinates {(\x,\y)};
				}		
				\begin{scope}[xshift=-0.15cm, yshift=-0.15cm]
					\draw (\x+0.03,0.6)--(\x+0.3-0.03,0.6)--(\x+0.3-0.03,1.2)--(\x+0.03,1.2)--(\x+0.03,0.6);
				\end{scope}
		}
			\draw[thick,->] (0.60,1.5)--(0.60,1.2);
		\end{scope}
		\begin{scope}[yshift=-1.8cm]
			\draw (-0.5,0) node[] {$t$};
			\draw (-0.5,0.3) node[] {$t+1$};
			\draw (-0.5,0.6) node[] {$t+2$};
			\foreach \x in {0,0.3,...,1.5}{
				\draw[red] plot [only marks, mark size=2.5, mark=*] coordinates {(\x,0.9)};
				\foreach \y in {0,0.3,0.6}{
					\draw[black] plot [only marks, mark size=2.5, mark=*] coordinates {(\x,\y)};
				}	
			}
			\draw[thick,<-] (1.5,0.5)--(1.75,0.5);
		\end{scope}
	\end{tikzpicture}
	\caption{The initialized system qubits(red) are sequentially swapped with the qubits of the bath-sink composite. In the end, the system-bath composite stores a slab of width $w-1$ and the system qubits store the missing row of the slab. \label{fig:swaps}}
\end{figure}
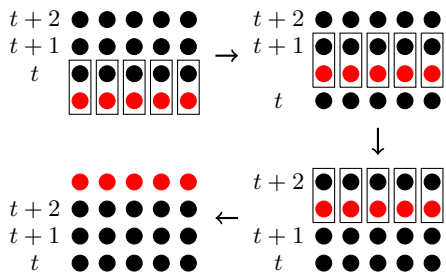

Intuitively, the action of the circuit describes the following physical process. At each time $t$, the bath-sink composite stores the reduced density matrix of the ground state over a slab of width $w$ which is comparable to the correlation length; the slab ranges from $y=t$ to $y=t+w$. The $t$-th row is swapped with the system qubits at time $t$, and then the $(k)$th row is swapped with the $(k+1)$th row from $k=t$ to $k=t+w-1$. At this point, the system qubit at time $t$ stores the qubits of the ground state on the $(t)$th row and the bath-sink composite stores a slab of width $w-1$; see FIG.\ref{fig:swaps}. Then, another finite-depth circuit is applied, which extends a slab of width $w-1$ to a slab of width $w$; the former ranges from $y=t+1$ to $y=t+w$ and the latter ranges from $y=t+1$ to $y=t+w+1$.

Only in the last step do we need the special property of the physical states, specifically, ground states of 2D systems with a mass gap. For certain exactly solvable models, circuits that achieve this goal have been already discussed in the literature\cite{Dennis2001,Milsted2016}. These circuits are constructed in such a way that a ground state of a system can be created by applying a low-depth circuit on the boundary of a smaller system. 

A more general argument involves the effective field theory description of gapped systems, the topological quantum field theory\cite{Witten1989}. The theory predicts that the entanglement entropy obeys the area law:
\begin{equation}
	S(A)=\alpha \ell - \gamma + \cdots, \label{eq:TEE}
\end{equation}
where $\ell$ is the perimeter of $A$, $\alpha$ is a non-universal constant, $\gamma$ is the topological entanglement entropy, and the ellipsis is the subcorrection term that vanishes in the $\ell \to \infty$ limit\cite{Kitaev2006}. As we show in the Supplemental Material,  Eq.\ref{eq:TEE} implies the existence of a local, finite-depth quantum circuit that extends the slab. 

The existence of such a circuit is an exact statement for certain solvable models, such as quantum double\cite{Kitaev2003} or the string-net models\cite{Levin2005}. Away from the fixed point, we expect the depth of the circuit to be determined by the correlation length of the system, which would also determine the number of necessary bath qubits. We expect the ratio between the number of bath qubits and system qubits to determine the largest correlation length the simulator can probe.

All the steps outlined above only involve finite-depth local quantum circuits. Therefore, for gapped systems, it suffices to only consider circuits of such form.\footnote{The circuit that creates the \emph{entire state}, as opposed to a single row, is not finite-depth; it scales linearly with the system size. This is unavoidable due to the existence of topological quantum order\cite{Bravyi2006,Koenig2014}.}

\textit{Local stability---}
While the circuits within each time steps remain local and bounded-depth, the circuit depth in its entirety scales linearly with $\ell_y$. In principle, noise can accumulate in time, and render the outcome of the simulation useless. For a general observable, this problem is surely unavoidable. However, local observables can be free of such problems; if the dissipative map induced on the bath qubits quickly equilibrates local observables, then expectation values of local observables can be perturbed by at most $O(\epsilon)$, independent of system size.

The core argument can be summarized as follows. Let us first formulate Eq.\ref{eq:fcs} in the Heisenberg picture. Since the measurement is only performed on the system qubits, the observable that the experimentalist measures is initially supported on the system qubits. However, since the dual of $\Phi_t$ maps operators on $S_tBS'$ to operators on $BS'$, such observables are mapped into an operator on $BS'$ and then evolve by a sequence of channels that are restricted to $BS'$. At every time step, the underlying circuits are bounded-depth and local, and as such, the support of the operator expands by at most by a constant amount. If local observables equilibrate quickly,  the operator will equilibrate to its steady state value before the support expands too much. Consequently, the circuit elements that appear after the equilibration becomes immaterial. A more  detailed analysis shall appear elsewhere\cite{Kim2017}.

\textit{Quantum versus classical variational methods---}
It is interesting to compare our method to the existing numerical methods that use a classical computer. The energy \emph{per site}, which is the figure of merit in comparing different numerical methods, is limited by some fixed precision. In our proposal, this precision is determined by the noise rate; this is because every local observable is perturbed by at most $O(\epsilon)$. In the existing numerical methods, it is determined by the numerical precision and the details of the algorithm. Furthermore, both methods are variational, because the energy is estimated from a valid quantum mechanical state.

The advantage of our method is that estimation of local observable can be done efficiently by design; see Eq.\ref{eq:state_definition}. Specifically, assuming that the circuit within each time steps are depth $D$, it takes time at most $O(\ell_y D/\delta^2)$ to estimate a local observable up to a precision $\delta$. In the existing numerical methods in dimensions higher than one, the same task scales as a high power of the variational parameter, or ought to be done approximately; see Ref.\cite{Evenbly2009,Verstraete2009} for the detail. On the other hand, unlike in the classical variational methods, commonly used subroutines such as singular value decomposition\cite{Evenbly2009,Verstraete2009} cannot be used in our method. To summarize, there are less options for energy minimization for our method, but the local expectation values can be estimated comparatively more efficiently. 

\textit{Resource estimate---}
As explained above, the simulation time depends on two factors: time for estimating the energy per site and the number of iterations till convergence. The latter depends on the detail of the model, but the former is determined by the parameters of the device. These parameter are $\tau$, time for executing the local gates,  $D$, the circuit depth, $\ell_x$, the number of system qubits, and $\delta$, the statistical error. They determine the time $T$ for estimating the energy as follows:
\begin{equation}
	T= C\frac{\tau D}{\ell_x \delta^2},\label{eq:time_estimate}
\end{equation}
where $C$ is a numerical constant of order unity which depends on the model.

This estimate follows from two simple observations. First, fully evolving the circuit takes $\tau D \ell_y$ time, where $\ell_y$ is the length of the simulated system in the $y$ direction. Second, in order to estimate the energy per site up to a statistical error $\delta$, one needs $O(\frac{1}{\delta^2\ell_x\ell_y})$ samples. The prefactor for the number of samples is the prefactor that appears in Eq.\ref{eq:time_estimate}. The gate time of superconducting qubits ranges from $50\text{ ns}$ to $500 \text{ ns}$\cite{Devoret2013}. This translates into $T$ in the order of miliseconds for $\ell_x=20$ and $\epsilon=0.01$. For local observables, the time can be similarly estimated as $T=C\tau D \ell_y / \delta^2$.

\textit{Discussion---}
We proposed a method to simulate 2D quantum many-body systems from a 1D simulator. The method is universally applicable to any quasi-1D platform equipped with local gates and measurements, and it also drastically reduces the number of qubits necessary for the simulation. A state whose energy per site is $O(\epsilon)$ away from the ground state can be prepared. An analogous conclusion holds in higher dimensions as well, e.g., for simulating three-dimensional quantum many-body systems from 2D simulators.

An important issue is the number of iterations of the variational method till convergence. The number will depend on the details of the model and the update rule. It is difficult to predict a general trend, because we should not expect the method to converge in systems that exhibit the spin glass phase.

The issue of local stability may be studied in quantum circuits\cite{Dennis2001,Milsted2016} for which the representation of the 2D system as a history of a 1D system becomes exact. The techniques in \cite{Cubitt2013,Lucia2014} play an important role. This analysis shall appear elsewhere\cite{Kim2017}.

There are two interesting future directions. First, it will be interesting to compare the reported energy of the simulation to that of the state-of-the-art numerical methods. Since the energy obtained from the simulation is guaranteed to be coming from a valid quantum mechanical state, they can be compared toe-to-toe. The advantage of our method is the low computational overhead in estimating the expectation values of local observables. The disadvantage would be the lack of options for energy minimization and a precision limited by the noise rate $\epsilon$, which is in the order of $10^{-2}$ in the leading architectures.

Second, assuming translation invariance, in the $\ell_y\to \infty$ limit, the state of the bath ought to encode all the properties of the 2D phase. Therefore, the steady state of the bath must possess a certain order that is inherited from the order of the 2D phase. The process of minimizing the energy in our context can be thus viewed as a process of preparing a one-dimensional phase of matter that is stabilized by a feedback control. Understanding their order parameters, as well as the phase transition between them, is left for future work. 

\textit{Acknowledgments}: I thank Kristan Temme, Jay Gambetta, Jens Eisert, Tobias Osborne, Soonwon Choi, and Hannes Pichler for helpful discussions.

\bibliography{bib}

\begin{thebibliography}{55}%
\makeatletter
\providecommand \@ifxundefined [1]{%
 \@ifx{#1\undefined}
}%
\providecommand \@ifnum [1]{%
 \ifnum #1\expandafter \@firstoftwo
 \else \expandafter \@secondoftwo
 \fi
}%
\providecommand \@ifx [1]{%
 \ifx #1\expandafter \@firstoftwo
 \else \expandafter \@secondoftwo
 \fi
}%
\providecommand \natexlab [1]{#1}%
\providecommand \enquote  [1]{``#1''}%
\providecommand \bibnamefont  [1]{#1}%
\providecommand \bibfnamefont [1]{#1}%
\providecommand \citenamefont [1]{#1}%
\providecommand \href@noop [0]{\@secondoftwo}%
\providecommand \href [0]{\begingroup \@sanitize@url \@href}%
\providecommand \@href[1]{\@@startlink{#1}\@@href}%
\providecommand \@@href[1]{\endgroup#1\@@endlink}%
\providecommand \@sanitize@url [0]{\catcode `\\12\catcode `\$12\catcode
  `\&12\catcode `\#12\catcode `\^12\catcode `\_12\catcode `\%12\relax}%
\providecommand \@@startlink[1]{}%
\providecommand \@@endlink[0]{}%
\providecommand \url  [0]{\begingroup\@sanitize@url \@url }%
\providecommand \@url [1]{\endgroup\@href {#1}{\urlprefix }}%
\providecommand \urlprefix  [0]{URL }%
\providecommand \Eprint [0]{\href }%
\providecommand \doibase [0]{http://dx.doi.org/}%
\providecommand \selectlanguage [0]{\@gobble}%
\providecommand \bibinfo  [0]{\@secondoftwo}%
\providecommand \bibfield  [0]{\@secondoftwo}%
\providecommand \translation [1]{[#1]}%
\providecommand \BibitemOpen [0]{}%
\providecommand \bibitemStop [0]{}%
\providecommand \bibitemNoStop [0]{.\EOS\space}%
\providecommand \EOS [0]{\spacefactor3000\relax}%
\providecommand \BibitemShut  [1]{\csname bibitem#1\endcsname}%
\let\auto@bib@innerbib\@empty
\bibitem [{\citenamefont {Turing}(1936)}]{Turing1936}%
  \BibitemOpen
  \bibfield  {author} {\bibinfo {author} {\bibfnamefont {A.~M.}\ \bibnamefont
  {Turing}},\ }\href@noop {} {\bibfield  {journal} {\bibinfo  {journal} {Proc.
  of the London Mathematical Society}\ }\textbf {\bibinfo {volume} {42}}
  (\bibinfo {year} {1936})}\BibitemShut {NoStop}%
\bibitem [{\citenamefont {Bush}\ \emph {et~al.}(1927)\citenamefont {Bush},
  \citenamefont {Gage},\ and\ \citenamefont {Stewart}}]{Bush1927}%
  \BibitemOpen
  \bibfield  {author} {\bibinfo {author} {\bibfnamefont {V.}~\bibnamefont
  {Bush}}, \bibinfo {author} {\bibfnamefont {F.}~\bibnamefont {Gage}}, \ and\
  \bibinfo {author} {\bibfnamefont {H.}~\bibnamefont {Stewart}},\ }\href
  {\doibase http://dx.doi.org/10.1016/S0016-0032(27)90097-0} {\bibfield
  {journal} {\bibinfo  {journal} {J. Franklin Inst.}\ }\textbf {\bibinfo
  {volume} {203}},\ \bibinfo {pages} {63 } (\bibinfo {year}
  {1927})}\BibitemShut {NoStop}%
\bibitem [{\citenamefont {Bush}(1931)}]{BUSH1931}%
  \BibitemOpen
  \bibfield  {author} {\bibinfo {author} {\bibfnamefont {V.}~\bibnamefont
  {Bush}},\ }\href {\doibase http://dx.doi.org/10.1016/S0016-0032(31)90616-9}
  {\bibfield  {journal} {\bibinfo  {journal} {J. Franklin Inst.}\ }\textbf
  {\bibinfo {volume} {212}},\ \bibinfo {pages} {447 } (\bibinfo {year}
  {1931})}\BibitemShut {NoStop}%
\bibitem [{\citenamefont {Shor}(1997)}]{Shor1997}%
  \BibitemOpen
  \bibfield  {author} {\bibinfo {author} {\bibfnamefont {P.~W.}\ \bibnamefont
  {Shor}},\ }\href@noop {} {\bibfield  {journal} {\bibinfo  {journal} {SIAM J.
  Comput.,}\ }\textbf {\bibinfo {volume} {26}},\ \bibinfo {pages} {1484}
  (\bibinfo {year} {1997})},\ \Eprint {http://arxiv.org/abs/quant-ph/9508027v2}
  {quant-ph/9508027v2} \BibitemShut {NoStop}%
\bibitem [{\citenamefont {Grover}(1996)}]{Grover1996}%
  \BibitemOpen
  \bibfield  {author} {\bibinfo {author} {\bibfnamefont {L.~K.}\ \bibnamefont
  {Grover}},\ }\href@noop {} {\bibfield  {journal} {\bibinfo  {journal}
  {Proceedings, 28th ACM Symposium on Theory of Computation}\ ,\ \bibinfo
  {pages} {212}} (\bibinfo {year} {1996})},\ \Eprint
  {http://arxiv.org/abs/quant-ph/9605043v3} {quant-ph/9605043v3} \BibitemShut
  {NoStop}%
\bibitem [{\citenamefont {Lloyd}(1996)}]{Lloyd1996}%
  \BibitemOpen
  \bibfield  {author} {\bibinfo {author} {\bibfnamefont {S.}~\bibnamefont
  {Lloyd}},\ }\href {\doibase 10.1126/science.273.5278.1073} {\bibfield
  {journal} {\bibinfo  {journal} {Science}\ }\textbf {\bibinfo {volume}
  {273}},\ \bibinfo {pages} {1073} (\bibinfo {year} {1996})}\BibitemShut
  {NoStop}%
\bibitem [{\citenamefont {Feynman}(1982)}]{Feynman1982}%
  \BibitemOpen
  \bibfield  {author} {\bibinfo {author} {\bibfnamefont {R.~P.}\ \bibnamefont
  {Feynman}},\ }\href {\doibase 10.1007/BF02650179} {\bibfield  {journal}
  {\bibinfo  {journal} {Int. J. Theor. Phys.}\ }\textbf {\bibinfo {volume}
  {21}},\ \bibinfo {pages} {467} (\bibinfo {year} {1982})}\BibitemShut
  {NoStop}%
\bibitem [{\citenamefont {Kim}\ \emph {et~al.}(2010)\citenamefont {Kim},
  \citenamefont {Chang}, \citenamefont {Korenblit}, \citenamefont {Islam},
  \citenamefont {Edwards}, \citenamefont {Freericks}, \citenamefont {Lin},
  \citenamefont {Duan},\ and\ \citenamefont {Monroe}}]{Kim2010}%
  \BibitemOpen
  \bibfield  {author} {\bibinfo {author} {\bibfnamefont {K.}~\bibnamefont
  {Kim}}, \bibinfo {author} {\bibfnamefont {M.-S.}\ \bibnamefont {Chang}},
  \bibinfo {author} {\bibfnamefont {S.}~\bibnamefont {Korenblit}}, \bibinfo
  {author} {\bibfnamefont {R.}~\bibnamefont {Islam}}, \bibinfo {author}
  {\bibfnamefont {E.~E.}\ \bibnamefont {Edwards}}, \bibinfo {author}
  {\bibfnamefont {J.~K.}\ \bibnamefont {Freericks}}, \bibinfo {author}
  {\bibfnamefont {G.-D.}\ \bibnamefont {Lin}}, \bibinfo {author} {\bibfnamefont
  {L.-M.}\ \bibnamefont {Duan}}, \ and\ \bibinfo {author} {\bibfnamefont
  {C.}~\bibnamefont {Monroe}},\ }\href {\doibase doi:10.1038/nature09071}
  {\bibfield  {journal} {\bibinfo  {journal} {Nature}\ }\textbf {\bibinfo
  {volume} {465}},\ \bibinfo {pages} {590} (\bibinfo {year}
  {2010})}\BibitemShut {NoStop}%
\bibitem [{\citenamefont {Simon}\ \emph {et~al.}(2011)\citenamefont {Simon},
  \citenamefont {Bakr}, \citenamefont {Ma}, \citenamefont {Tai}, \citenamefont
  {Preiss},\ and\ \citenamefont {Greiner}}]{Simon2011}%
  \BibitemOpen
  \bibfield  {author} {\bibinfo {author} {\bibfnamefont {J.}~\bibnamefont
  {Simon}}, \bibinfo {author} {\bibfnamefont {W.~S.}\ \bibnamefont {Bakr}},
  \bibinfo {author} {\bibfnamefont {R.}~\bibnamefont {Ma}}, \bibinfo {author}
  {\bibfnamefont {M.~E.}\ \bibnamefont {Tai}}, \bibinfo {author} {\bibfnamefont
  {P.~M.}\ \bibnamefont {Preiss}}, \ and\ \bibinfo {author} {\bibfnamefont
  {M.}~\bibnamefont {Greiner}},\ }\href@noop {} {\bibfield  {journal} {\bibinfo
   {journal} {Nature}\ }\textbf {\bibinfo {volume} {472}},\ \bibinfo {pages}
  {307} (\bibinfo {year} {2011})}\BibitemShut {NoStop}%
\bibitem [{\citenamefont {Trotzky}\ \emph {et~al.}(2012)\citenamefont
  {Trotzky}, \citenamefont {Chen}, \citenamefont {Flesch}, \citenamefont
  {McCulloch}, \citenamefont {Schollw\"ock}, \citenamefont {Eisert},\ and\
  \citenamefont {Bloch}}]{Trotzky2012}%
  \BibitemOpen
  \bibfield  {author} {\bibinfo {author} {\bibfnamefont {S.}~\bibnamefont
  {Trotzky}}, \bibinfo {author} {\bibfnamefont {Y.-A.}\ \bibnamefont {Chen}},
  \bibinfo {author} {\bibfnamefont {A.}~\bibnamefont {Flesch}}, \bibinfo
  {author} {\bibfnamefont {I.~P.}\ \bibnamefont {McCulloch}}, \bibinfo {author}
  {\bibfnamefont {U.}~\bibnamefont {Schollw\"ock}}, \bibinfo {author}
  {\bibfnamefont {J.}~\bibnamefont {Eisert}}, \ and\ \bibinfo {author}
  {\bibfnamefont {I.}~\bibnamefont {Bloch}},\ }\href {\doibase
  10.1038/nphys2232} {\bibfield  {journal} {\bibinfo  {journal} {Nature
  Physics}\ }\textbf {\bibinfo {volume} {8}},\ \bibinfo {pages} {325} (\bibinfo
  {year} {2012})},\ \Eprint {http://arxiv.org/abs/1101.2659v1} {1101.2659v1}
  \BibitemShut {NoStop}%
\bibitem [{\citenamefont {Greif}\ \emph {et~al.}(2013)\citenamefont {Greif},
  \citenamefont {Uehlinger}, \citenamefont {Jotzu}, \citenamefont {Tarruell},\
  and\ \citenamefont {Esslinger}}]{Greif2013}%
  \BibitemOpen
  \bibfield  {author} {\bibinfo {author} {\bibfnamefont {D.}~\bibnamefont
  {Greif}}, \bibinfo {author} {\bibfnamefont {T.}~\bibnamefont {Uehlinger}},
  \bibinfo {author} {\bibfnamefont {G.}~\bibnamefont {Jotzu}}, \bibinfo
  {author} {\bibfnamefont {L.}~\bibnamefont {Tarruell}}, \ and\ \bibinfo
  {author} {\bibfnamefont {T.}~\bibnamefont {Esslinger}},\ }\href {\doibase
  10.1126/science.1236362} {\bibfield  {journal} {\bibinfo  {journal}
  {Science}\ }\textbf {\bibinfo {volume} {340}},\ \bibinfo {pages} {1307}
  (\bibinfo {year} {2013})},\ \Eprint {http://arxiv.org/abs/1212.2634v2}
  {1212.2634v2} \BibitemShut {NoStop}%
\bibitem [{\citenamefont {Hart}\ \emph {et~al.}(2015)\citenamefont {Hart},
  \citenamefont {Duarte}, \citenamefont {Yang}, \citenamefont {Liu},
  \citenamefont {Paiva}, \citenamefont {Khatami}, \citenamefont {Scalettar},
  \citenamefont {Trivedi}, \citenamefont {Huse},\ and\ \citenamefont
  {Hulet}}]{Hart2015}%
  \BibitemOpen
  \bibfield  {author} {\bibinfo {author} {\bibfnamefont {R.~A.}\ \bibnamefont
  {Hart}}, \bibinfo {author} {\bibfnamefont {P.~M.}\ \bibnamefont {Duarte}},
  \bibinfo {author} {\bibfnamefont {T.-L.}\ \bibnamefont {Yang}}, \bibinfo
  {author} {\bibfnamefont {X.}~\bibnamefont {Liu}}, \bibinfo {author}
  {\bibfnamefont {T.}~\bibnamefont {Paiva}}, \bibinfo {author} {\bibfnamefont
  {E.}~\bibnamefont {Khatami}}, \bibinfo {author} {\bibfnamefont {R.~T.}\
  \bibnamefont {Scalettar}}, \bibinfo {author} {\bibfnamefont {N.}~\bibnamefont
  {Trivedi}}, \bibinfo {author} {\bibfnamefont {D.~A.}\ \bibnamefont {Huse}}, \
  and\ \bibinfo {author} {\bibfnamefont {R.~G.}\ \bibnamefont {Hulet}},\ }\href
  {\doibase 10.1038/nature14223} {\bibfield  {journal} {\bibinfo  {journal}
  {Nature}\ }\textbf {\bibinfo {volume} {519}},\ \bibinfo {pages} {211}
  (\bibinfo {year} {2015})},\ \Eprint {http://arxiv.org/abs/1407.5932v2}
  {1407.5932v2} \BibitemShut {NoStop}%
\bibitem [{\citenamefont {{Hubbard}}(1963)}]{Hubbard1963}%
  \BibitemOpen
  \bibfield  {author} {\bibinfo {author} {\bibfnamefont {J.}~\bibnamefont
  {{Hubbard}}},\ }\href {\doibase 10.1098/rspa.1963.0204} {\bibfield  {journal}
  {\bibinfo  {journal} {Proc. Roy. Soc. (London), Ser. A}\ }\textbf {\bibinfo
  {volume} {276}},\ \bibinfo {pages} {238} (\bibinfo {year}
  {1963})}\BibitemShut {NoStop}%
\bibitem [{\citenamefont {Anderson}(1987)}]{Anderson1987}%
  \BibitemOpen
  \bibfield  {author} {\bibinfo {author} {\bibfnamefont {P.~W.}\ \bibnamefont
  {Anderson}},\ }\href@noop {} {\bibfield  {journal} {\bibinfo  {journal}
  {Science}\ }\textbf {\bibinfo {volume} {235}},\ \bibinfo {pages} {1196}
  (\bibinfo {year} {1987})}\BibitemShut {NoStop}%
\bibitem [{\citenamefont {White}(1992)}]{White1992}%
  \BibitemOpen
  \bibfield  {author} {\bibinfo {author} {\bibfnamefont {S.~R.}\ \bibnamefont
  {White}},\ }\href {\doibase 10.1103/PhysRevLett.69.2863} {\bibfield
  {journal} {\bibinfo  {journal} {Phys. Rev. Lett.}\ }\textbf {\bibinfo
  {volume} {69}},\ \bibinfo {pages} {2863} (\bibinfo {year}
  {1992})}\BibitemShut {NoStop}%
\bibitem [{\citenamefont {Verstraete}\ and\ \citenamefont
  {Cirac}(2004)}]{Verstraete2004a}%
  \BibitemOpen
  \bibfield  {author} {\bibinfo {author} {\bibfnamefont {F.}~\bibnamefont
  {Verstraete}}\ and\ \bibinfo {author} {\bibfnamefont {J.~I.}\ \bibnamefont
  {Cirac}},\ }\href@noop {} {\  (\bibinfo {year} {2004})},\ \Eprint
  {http://arxiv.org/abs/cond-mat/0407066v1} {cond-mat/0407066v1} \BibitemShut
  {NoStop}%
\bibitem [{\citenamefont {Vidal}(2008)}]{Vidal2008}%
  \BibitemOpen
  \bibfield  {author} {\bibinfo {author} {\bibfnamefont {G.}~\bibnamefont
  {Vidal}},\ }\href {\doibase 10.1103/PhysRevLett.101.110501} {\bibfield
  {journal} {\bibinfo  {journal} {Phys. Rev. Lett.}\ }\textbf {\bibinfo
  {volume} {101}},\ \bibinfo {pages} {110501} (\bibinfo {year}
  {2008})}\BibitemShut {NoStop}%
\bibitem [{\citenamefont {McKay}\ and\ \citenamefont
  {DeMarco}(2011)}]{McKay2011}%
  \BibitemOpen
  \bibfield  {author} {\bibinfo {author} {\bibfnamefont {D.~C.}\ \bibnamefont
  {McKay}}\ and\ \bibinfo {author} {\bibfnamefont {B.}~\bibnamefont
  {DeMarco}},\ }\href {http://stacks.iop.org/0034-4885/74/i=5/a=054401}
  {\bibfield  {journal} {\bibinfo  {journal} {Rep. Prog. Phys.}\ }\textbf
  {\bibinfo {volume} {74}},\ \bibinfo {pages} {054401} (\bibinfo {year}
  {2011})}\BibitemShut {NoStop}%
\bibitem [{\citenamefont {Lanyon}\ \emph {et~al.}(2011)\citenamefont {Lanyon},
  \citenamefont {Hempel}, \citenamefont {Nigg}, \citenamefont {M\"uller},
  \citenamefont {Gerritsma}, \citenamefont {Z\"ahringer}, \citenamefont
  {Schindler}, \citenamefont {Barreiro}, \citenamefont {Rambach}, \citenamefont
  {Kirchmair}, \citenamefont {Hennrich}, \citenamefont {Zoller}, \citenamefont
  {Blatt},\ and\ \citenamefont {Roos}}]{Lanyon2011}%
  \BibitemOpen
  \bibfield  {author} {\bibinfo {author} {\bibfnamefont {B.~P.}\ \bibnamefont
  {Lanyon}}, \bibinfo {author} {\bibfnamefont {C.}~\bibnamefont {Hempel}},
  \bibinfo {author} {\bibfnamefont {D.}~\bibnamefont {Nigg}}, \bibinfo {author}
  {\bibfnamefont {M.}~\bibnamefont {M\"uller}}, \bibinfo {author}
  {\bibfnamefont {R.}~\bibnamefont {Gerritsma}}, \bibinfo {author}
  {\bibfnamefont {F.}~\bibnamefont {Z\"ahringer}}, \bibinfo {author}
  {\bibfnamefont {P.}~\bibnamefont {Schindler}}, \bibinfo {author}
  {\bibfnamefont {J.~T.}\ \bibnamefont {Barreiro}}, \bibinfo {author}
  {\bibfnamefont {M.}~\bibnamefont {Rambach}}, \bibinfo {author} {\bibfnamefont
  {G.}~\bibnamefont {Kirchmair}}, \bibinfo {author} {\bibfnamefont
  {M.}~\bibnamefont {Hennrich}}, \bibinfo {author} {\bibfnamefont
  {P.}~\bibnamefont {Zoller}}, \bibinfo {author} {\bibfnamefont
  {R.}~\bibnamefont {Blatt}}, \ and\ \bibinfo {author} {\bibfnamefont {C.~F.}\
  \bibnamefont {Roos}},\ }\href {\doibase 10.1126/science.1208001} {\bibfield
  {journal} {\bibinfo  {journal} {Science}\ }\textbf {\bibinfo {volume}
  {334}},\ \bibinfo {pages} {57} (\bibinfo {year} {2011})},\ \Eprint
  {http://arxiv.org/abs/1109.1512v2} {1109.1512v2} \BibitemShut {NoStop}%
\bibitem [{\citenamefont {Blatt}\ and\ \citenamefont {Roos}(2012)}]{Blatt2012}%
  \BibitemOpen
  \bibfield  {author} {\bibinfo {author} {\bibfnamefont {R.}~\bibnamefont
  {Blatt}}\ and\ \bibinfo {author} {\bibfnamefont {C.~F.}\ \bibnamefont
  {Roos}},\ }\href@noop {} {\bibfield  {journal} {\bibinfo  {journal} {Nature
  Physics}\ }\textbf {\bibinfo {volume} {8}},\ \bibinfo {pages} {277} (\bibinfo
  {year} {2012})}\BibitemShut {NoStop}%
\bibitem [{\citenamefont {Islam}\ \emph {et~al.}(2013)\citenamefont {Islam},
  \citenamefont {Senko}, \citenamefont {Campbell}, \citenamefont {Korenblit},
  \citenamefont {Smith}, \citenamefont {Lee}, \citenamefont {Edwards},
  \citenamefont {Wang}, \citenamefont {Freericks},\ and\ \citenamefont
  {Monroe}}]{Islam2012}%
  \BibitemOpen
  \bibfield  {author} {\bibinfo {author} {\bibfnamefont {R.}~\bibnamefont
  {Islam}}, \bibinfo {author} {\bibfnamefont {C.}~\bibnamefont {Senko}},
  \bibinfo {author} {\bibfnamefont {W.~C.}\ \bibnamefont {Campbell}}, \bibinfo
  {author} {\bibfnamefont {S.}~\bibnamefont {Korenblit}}, \bibinfo {author}
  {\bibfnamefont {J.}~\bibnamefont {Smith}}, \bibinfo {author} {\bibfnamefont
  {A.}~\bibnamefont {Lee}}, \bibinfo {author} {\bibfnamefont {E.~E.}\
  \bibnamefont {Edwards}}, \bibinfo {author} {\bibfnamefont {C.~C.~J.}\
  \bibnamefont {Wang}}, \bibinfo {author} {\bibfnamefont {J.~K.}\ \bibnamefont
  {Freericks}}, \ and\ \bibinfo {author} {\bibfnamefont {C.}~\bibnamefont
  {Monroe}},\ }\href {\doibase 10.1126/science.1232296} {\bibfield  {journal}
  {\bibinfo  {journal} {Science}\ }\textbf {\bibinfo {volume} {340}},\ \bibinfo
  {pages} {583} (\bibinfo {year} {2013})},\ \Eprint
  {http://arxiv.org/abs/1210.0142v1} {1210.0142v1} \BibitemShut {NoStop}%
\bibitem [{\citenamefont {Houck}\ \emph {et~al.}(2012)\citenamefont {Houck},
  \citenamefont {T\"ureci},\ and\ \citenamefont {Koch}}]{Houck2012}%
  \BibitemOpen
  \bibfield  {author} {\bibinfo {author} {\bibfnamefont {A.~A.}\ \bibnamefont
  {Houck}}, \bibinfo {author} {\bibfnamefont {H.~E.}\ \bibnamefont {T\"ureci}},
  \ and\ \bibinfo {author} {\bibfnamefont {J.}~\bibnamefont {Koch}},\
  }\href@noop {} {\bibfield  {journal} {\bibinfo  {journal} {Nature Physics}\
  }\textbf {\bibinfo {volume} {8}},\ \bibinfo {pages} {292} (\bibinfo {year}
  {2012})}\BibitemShut {NoStop}%
\bibitem [{\citenamefont {Barends}\ \emph {et~al.}(2015)\citenamefont
  {Barends}, \citenamefont {Lamata}, \citenamefont {Kelly}, \citenamefont
  {García-Alvarez}, \citenamefont {Fowler}, \citenamefont {Megrant},
  \citenamefont {Jeffrey}, \citenamefont {White}, \citenamefont {Sank},
  \citenamefont {Mutus}, \citenamefont {Campbell}, \citenamefont {Chen},
  \citenamefont {Chen}, \citenamefont {Chiaro}, \citenamefont {Dunsworth},
  \citenamefont {Hoi}, \citenamefont {Neill}, \citenamefont {O'Malley},
  \citenamefont {Quintana}, \citenamefont {Roushan}, \citenamefont
  {Vainsencher}, \citenamefont {Wenner}, \citenamefont {Solano},\ and\
  \citenamefont {Martinis}}]{Barends2015a}%
  \BibitemOpen
  \bibfield  {author} {\bibinfo {author} {\bibfnamefont {R.}~\bibnamefont
  {Barends}}, \bibinfo {author} {\bibfnamefont {L.}~\bibnamefont {Lamata}},
  \bibinfo {author} {\bibfnamefont {J.}~\bibnamefont {Kelly}}, \bibinfo
  {author} {\bibfnamefont {L.}~\bibnamefont {García-Alvarez}}, \bibinfo
  {author} {\bibfnamefont {A.~G.}\ \bibnamefont {Fowler}}, \bibinfo {author}
  {\bibfnamefont {A.}~\bibnamefont {Megrant}}, \bibinfo {author} {\bibfnamefont
  {E.}~\bibnamefont {Jeffrey}}, \bibinfo {author} {\bibfnamefont {T.~C.}\
  \bibnamefont {White}}, \bibinfo {author} {\bibfnamefont {D.}~\bibnamefont
  {Sank}}, \bibinfo {author} {\bibfnamefont {J.~Y.}\ \bibnamefont {Mutus}},
  \bibinfo {author} {\bibfnamefont {B.}~\bibnamefont {Campbell}}, \bibinfo
  {author} {\bibfnamefont {Y.}~\bibnamefont {Chen}}, \bibinfo {author}
  {\bibfnamefont {Z.}~\bibnamefont {Chen}}, \bibinfo {author} {\bibfnamefont
  {B.}~\bibnamefont {Chiaro}}, \bibinfo {author} {\bibfnamefont
  {A.}~\bibnamefont {Dunsworth}}, \bibinfo {author} {\bibfnamefont {I.~C.}\
  \bibnamefont {Hoi}}, \bibinfo {author} {\bibfnamefont {C.}~\bibnamefont
  {Neill}}, \bibinfo {author} {\bibfnamefont {P.~J.~J.}\ \bibnamefont
  {O'Malley}}, \bibinfo {author} {\bibfnamefont {C.}~\bibnamefont {Quintana}},
  \bibinfo {author} {\bibfnamefont {P.}~\bibnamefont {Roushan}}, \bibinfo
  {author} {\bibfnamefont {A.}~\bibnamefont {Vainsencher}}, \bibinfo {author}
  {\bibfnamefont {J.}~\bibnamefont {Wenner}}, \bibinfo {author} {\bibfnamefont
  {E.}~\bibnamefont {Solano}}, \ and\ \bibinfo {author} {\bibfnamefont {J.~M.}\
  \bibnamefont {Martinis}},\ }\href {\doibase 10.1038/ncomms8654} {\bibfield
  {journal} {\bibinfo  {journal} {Nat. Comm.}\ }\textbf {\bibinfo {volume}
  {6}},\ \bibinfo {pages} {7654} (\bibinfo {year} {2015})},\ \Eprint
  {http://arxiv.org/abs/1501.07703v1} {1501.07703v1} \BibitemShut {NoStop}%
\bibitem [{\citenamefont {Barends}\ \emph {et~al.}(2016)\citenamefont
  {Barends}, \citenamefont {Shabani}, \citenamefont {Lamata}, \citenamefont
  {Kelly}, \citenamefont {Mezzacapo}, \citenamefont {Heras}, \citenamefont
  {Babbush}, \citenamefont {Fowler}, \citenamefont {Campbell}, \citenamefont
  {Chen}, \citenamefont {Chen}, \citenamefont {Chiaro}, \citenamefont
  {Dunsworth}, \citenamefont {Jeffrey}, \citenamefont {Lucero}, \citenamefont
  {Megrant}, \citenamefont {Mutus}, \citenamefont {Neeley}, \citenamefont
  {Neill}, \citenamefont {O'Malley}, \citenamefont {Quintana}, \citenamefont
  {Roushan}, \citenamefont {Sank}, \citenamefont {Vainsencher}, \citenamefont
  {Wenner}, \citenamefont {White}, \citenamefont {Solano}, \citenamefont
  {Neven},\ and\ \citenamefont {Martinis}}]{Barends2015}%
  \BibitemOpen
  \bibfield  {author} {\bibinfo {author} {\bibfnamefont {R.}~\bibnamefont
  {Barends}}, \bibinfo {author} {\bibfnamefont {A.}~\bibnamefont {Shabani}},
  \bibinfo {author} {\bibfnamefont {L.}~\bibnamefont {Lamata}}, \bibinfo
  {author} {\bibfnamefont {J.}~\bibnamefont {Kelly}}, \bibinfo {author}
  {\bibfnamefont {A.}~\bibnamefont {Mezzacapo}}, \bibinfo {author}
  {\bibfnamefont {U.~L.}\ \bibnamefont {Heras}}, \bibinfo {author}
  {\bibfnamefont {R.}~\bibnamefont {Babbush}}, \bibinfo {author} {\bibfnamefont
  {A.~G.}\ \bibnamefont {Fowler}}, \bibinfo {author} {\bibfnamefont
  {B.}~\bibnamefont {Campbell}}, \bibinfo {author} {\bibfnamefont
  {Y.}~\bibnamefont {Chen}}, \bibinfo {author} {\bibfnamefont {Z.}~\bibnamefont
  {Chen}}, \bibinfo {author} {\bibfnamefont {B.}~\bibnamefont {Chiaro}},
  \bibinfo {author} {\bibfnamefont {A.}~\bibnamefont {Dunsworth}}, \bibinfo
  {author} {\bibfnamefont {E.}~\bibnamefont {Jeffrey}}, \bibinfo {author}
  {\bibfnamefont {E.}~\bibnamefont {Lucero}}, \bibinfo {author} {\bibfnamefont
  {A.}~\bibnamefont {Megrant}}, \bibinfo {author} {\bibfnamefont {J.~Y.}\
  \bibnamefont {Mutus}}, \bibinfo {author} {\bibfnamefont {M.}~\bibnamefont
  {Neeley}}, \bibinfo {author} {\bibfnamefont {C.}~\bibnamefont {Neill}},
  \bibinfo {author} {\bibfnamefont {P.~J.~J.}\ \bibnamefont {O'Malley}},
  \bibinfo {author} {\bibfnamefont {C.}~\bibnamefont {Quintana}}, \bibinfo
  {author} {\bibfnamefont {P.}~\bibnamefont {Roushan}}, \bibinfo {author}
  {\bibfnamefont {D.}~\bibnamefont {Sank}}, \bibinfo {author} {\bibfnamefont
  {A.}~\bibnamefont {Vainsencher}}, \bibinfo {author} {\bibfnamefont
  {J.}~\bibnamefont {Wenner}}, \bibinfo {author} {\bibfnamefont {T.~C.}\
  \bibnamefont {White}}, \bibinfo {author} {\bibfnamefont {E.}~\bibnamefont
  {Solano}}, \bibinfo {author} {\bibfnamefont {H.}~\bibnamefont {Neven}}, \
  and\ \bibinfo {author} {\bibfnamefont {J.~M.}\ \bibnamefont {Martinis}},\
  }\href {\doibase 10.1038/nature17658} {\bibfield  {journal} {\bibinfo
  {journal} {Nature}\ }\textbf {\bibinfo {volume} {534}},\ \bibinfo {pages}
  {222} (\bibinfo {year} {2016})},\ \Eprint {http://arxiv.org/abs/1511.03316v1}
  {1511.03316v1} \BibitemShut {NoStop}%
\bibitem [{\citenamefont {Smith}\ \emph {et~al.}(2016)\citenamefont {Smith},
  \citenamefont {Lee}, \citenamefont {Richerme}, \citenamefont {Neyenhuis},
  \citenamefont {Hess}, \citenamefont {Hauke}, \citenamefont {Heyl},
  \citenamefont {Huse},\ and\ \citenamefont {Monroe}}]{Smith2015}%
  \BibitemOpen
  \bibfield  {author} {\bibinfo {author} {\bibfnamefont {J.}~\bibnamefont
  {Smith}}, \bibinfo {author} {\bibfnamefont {A.}~\bibnamefont {Lee}}, \bibinfo
  {author} {\bibfnamefont {P.}~\bibnamefont {Richerme}}, \bibinfo {author}
  {\bibfnamefont {B.}~\bibnamefont {Neyenhuis}}, \bibinfo {author}
  {\bibfnamefont {P.~W.}\ \bibnamefont {Hess}}, \bibinfo {author}
  {\bibfnamefont {P.}~\bibnamefont {Hauke}}, \bibinfo {author} {\bibfnamefont
  {M.}~\bibnamefont {Heyl}}, \bibinfo {author} {\bibfnamefont {D.~A.}\
  \bibnamefont {Huse}}, \ and\ \bibinfo {author} {\bibfnamefont
  {C.}~\bibnamefont {Monroe}},\ }\href {\doibase 10.1038/nphys3783} {\bibfield
  {journal} {\bibinfo  {journal} {Nature Physics}\ }\textbf {\bibinfo {volume}
  {12}},\ \bibinfo {pages} {907} (\bibinfo {year} {2016})},\ \Eprint
  {http://arxiv.org/abs/1508.07026v1} {1508.07026v1} \BibitemShut {NoStop}%
\bibitem [{\citenamefont {Hastings}(2007)}]{Hastings2007}%
  \BibitemOpen
  \bibfield  {author} {\bibinfo {author} {\bibfnamefont {M.~B.}\ \bibnamefont
  {Hastings}},\ }\href@noop {} {\bibfield  {journal} {\bibinfo  {journal}
  {JSTAT}\ } (\bibinfo {year} {2007})},\ \Eprint
  {http://arxiv.org/abs/0705.2024v3} {0705.2024v3} \BibitemShut {NoStop}%
\bibitem [{\citenamefont {Eisert}\ \emph {et~al.}(2010)\citenamefont {Eisert},
  \citenamefont {Cramer},\ and\ \citenamefont {Plenio}}]{Eisert2010}%
  \BibitemOpen
  \bibfield  {author} {\bibinfo {author} {\bibfnamefont {J.}~\bibnamefont
  {Eisert}}, \bibinfo {author} {\bibfnamefont {M.}~\bibnamefont {Cramer}}, \
  and\ \bibinfo {author} {\bibfnamefont {M.~B.}\ \bibnamefont {Plenio}},\
  }\href {\doibase 10.1103/RevModPhys.82.277} {\bibfield  {journal} {\bibinfo
  {journal} {Rev. Mod. Phys.}\ }\textbf {\bibinfo {volume} {82}},\ \bibinfo
  {pages} {277} (\bibinfo {year} {2010})}\BibitemShut {NoStop}%
\bibitem [{\citenamefont {Ge}\ and\ \citenamefont {Eisert}(2016)}]{Ge2014}%
  \BibitemOpen
  \bibfield  {author} {\bibinfo {author} {\bibfnamefont {Y.}~\bibnamefont
  {Ge}}\ and\ \bibinfo {author} {\bibfnamefont {J.}~\bibnamefont {Eisert}},\
  }\href {http://arxiv.org/abs/1411.2995} {\bibfield  {journal} {\bibinfo
  {journal} {New J. Phys.}\ }\textbf {\bibinfo {volume} {18}},\ \bibinfo
  {pages} {083026} (\bibinfo {year} {2016})},\ \Eprint
  {http://arxiv.org/abs/1411.2995} {arXiv:1411.2995} \BibitemShut {NoStop}%
\bibitem [{\citenamefont {Kitaev}\ and\ \citenamefont
  {Preskill}(2006)}]{Kitaev2006}%
  \BibitemOpen
  \bibfield  {author} {\bibinfo {author} {\bibfnamefont {A.}~\bibnamefont
  {Kitaev}}\ and\ \bibinfo {author} {\bibfnamefont {J.}~\bibnamefont
  {Preskill}},\ }\href {\doibase 10.1103/PhysRevLett.96.110404} {\bibfield
  {journal} {\bibinfo  {journal} {Phys. Rev. Lett.}\ }\textbf {\bibinfo
  {volume} {96}},\ \bibinfo {pages} {110404} (\bibinfo {year} {2006})},\
  \Eprint {http://arxiv.org/abs/hep-th/0510092v2} {hep-th/0510092v2}
  \BibitemShut {NoStop}%
\bibitem [{\citenamefont {Levin}\ and\ \citenamefont {Wen}(2006)}]{Levin2006}%
  \BibitemOpen
  \bibfield  {author} {\bibinfo {author} {\bibfnamefont {M.}~\bibnamefont
  {Levin}}\ and\ \bibinfo {author} {\bibfnamefont {X.-G.}\ \bibnamefont
  {Wen}},\ }\href {\doibase 10.1103/PhysRevLett.96.110405} {\bibfield
  {journal} {\bibinfo  {journal} {Phys. Rev. Lett.}\ }\textbf {\bibinfo
  {volume} {96}},\ \bibinfo {pages} {110405} (\bibinfo {year}
  {2006})}\BibitemShut {NoStop}%
\bibitem [{\citenamefont {Kim}(2014)}]{Kim2014}%
  \BibitemOpen
  \bibfield  {author} {\bibinfo {author} {\bibfnamefont {I.~H.}\ \bibnamefont
  {Kim}},\ }\href@noop {} {\  (\bibinfo {year} {2014})},\ \Eprint
  {http://arxiv.org/abs/1405.0137v1} {1405.0137v1} \BibitemShut {NoStop}%
\bibitem [{Note1()}]{Note1}%
  \BibitemOpen
  \bibinfo {note} {In systems with Goldstone modes, we expect the noise to
  effectively act as a symmetry-breaking term. The noise rate is likely to
  determine the mass.}\BibitemShut {Stop}%
\bibitem [{\citenamefont {Peruzzo}\ \emph {et~al.}(2014)\citenamefont
  {Peruzzo}, \citenamefont {McClean}, \citenamefont {Shadbolt}, \citenamefont
  {Yung}, \citenamefont {Zhou}, \citenamefont {Love}, \citenamefont
  {Aspuru-Guzik},\ and\ \citenamefont {O'Brien}}]{Peruzzo2013}%
  \BibitemOpen
  \bibfield  {author} {\bibinfo {author} {\bibfnamefont {A.}~\bibnamefont
  {Peruzzo}}, \bibinfo {author} {\bibfnamefont {J.}~\bibnamefont {McClean}},
  \bibinfo {author} {\bibfnamefont {P.}~\bibnamefont {Shadbolt}}, \bibinfo
  {author} {\bibfnamefont {M.-H.}\ \bibnamefont {Yung}}, \bibinfo {author}
  {\bibfnamefont {X.-Q.}\ \bibnamefont {Zhou}}, \bibinfo {author}
  {\bibfnamefont {P.~J.}\ \bibnamefont {Love}}, \bibinfo {author}
  {\bibfnamefont {A.}~\bibnamefont {Aspuru-Guzik}}, \ and\ \bibinfo {author}
  {\bibfnamefont {J.~L.}\ \bibnamefont {O'Brien}},\ }\href {\doibase
  10.1038/ncomms5213} {\bibfield  {journal} {\bibinfo  {journal} {Nature
  Communications}\ }\textbf {\bibinfo {volume} {5}} (\bibinfo {year} {2014}),\
  10.1038/ncomms5213},\ \Eprint {http://arxiv.org/abs/1304.3061v1}
  {1304.3061v1} \BibitemShut {NoStop}%
\bibitem [{\citenamefont {Osborne}\ \emph {et~al.}(2010)\citenamefont
  {Osborne}, \citenamefont {Eisert},\ and\ \citenamefont
  {Verstraete}}]{Osborne2010}%
  \BibitemOpen
  \bibfield  {author} {\bibinfo {author} {\bibfnamefont {T.~J.}\ \bibnamefont
  {Osborne}}, \bibinfo {author} {\bibfnamefont {J.}~\bibnamefont {Eisert}}, \
  and\ \bibinfo {author} {\bibfnamefont {F.}~\bibnamefont {Verstraete}},\
  }\href {\doibase 10.1103/PhysRevLett.105.260401} {\bibfield  {journal}
  {\bibinfo  {journal} {Phys. Rev. Lett.}\ }\textbf {\bibinfo {volume} {105}},\
  \bibinfo {pages} {260401} (\bibinfo {year} {2010})}\BibitemShut {NoStop}%
\bibitem [{\citenamefont {Barrett}\ \emph {et~al.}(2013)\citenamefont
  {Barrett}, \citenamefont {Hammerer}, \citenamefont {Harrison}, \citenamefont
  {Northup},\ and\ \citenamefont {Osborne}}]{Barrett2013}%
  \BibitemOpen
  \bibfield  {author} {\bibinfo {author} {\bibfnamefont {S.}~\bibnamefont
  {Barrett}}, \bibinfo {author} {\bibfnamefont {K.}~\bibnamefont {Hammerer}},
  \bibinfo {author} {\bibfnamefont {S.}~\bibnamefont {Harrison}}, \bibinfo
  {author} {\bibfnamefont {T.~E.}\ \bibnamefont {Northup}}, \ and\ \bibinfo
  {author} {\bibfnamefont {T.~J.}\ \bibnamefont {Osborne}},\ }\href {\doibase
  10.1103/PhysRevLett.110.090501} {\bibfield  {journal} {\bibinfo  {journal}
  {Phys. Rev. Lett.}\ }\textbf {\bibinfo {volume} {110}},\ \bibinfo {pages}
  {090501} (\bibinfo {year} {2013})}\BibitemShut {NoStop}%
\bibitem [{Note2()}]{Note2}%
  \BibitemOpen
  \bibinfo {note} {One can make the architecture to be truly 1D with the
  expense of allowing slightly nonlocal interaction, e.g.,
  next-nearest-neighbor interaction.}\BibitemShut {Stop}%
\bibitem [{\citenamefont {Nielsen}\ and\ \citenamefont
  {Chuang}(2011)}]{Nielsen2011}%
  \BibitemOpen
  \bibfield  {author} {\bibinfo {author} {\bibfnamefont {M.~A.}\ \bibnamefont
  {Nielsen}}\ and\ \bibinfo {author} {\bibfnamefont {I.~L.}\ \bibnamefont
  {Chuang}},\ }\href@noop {} {\emph {\bibinfo {title} {Quantum Computation and
  Quantum Information: 10th Anniversary Edition}}},\ \bibinfo {edition} {10th}\
  ed.\ (\bibinfo  {publisher} {Cambridge University Press},\ \bibinfo {address}
  {New York, NY, USA},\ \bibinfo {year} {2011})\BibitemShut {NoStop}%
\bibitem [{\citenamefont {Fannes}\ \emph {et~al.}(1992)\citenamefont {Fannes},
  \citenamefont {Nachtergaele},\ and\ \citenamefont {Werner}}]{Fannes1992}%
  \BibitemOpen
  \bibfield  {author} {\bibinfo {author} {\bibfnamefont {M.}~\bibnamefont
  {Fannes}}, \bibinfo {author} {\bibfnamefont {B.}~\bibnamefont
  {Nachtergaele}}, \ and\ \bibinfo {author} {\bibfnamefont {R.~F.}\
  \bibnamefont {Werner}},\ }\href
  {http://projecteuclid.org/euclid.cmp/1104249404} {\bibfield  {journal}
  {\bibinfo  {journal} {Comm. Math. Phys.}\ }\textbf {\bibinfo {volume}
  {144}},\ \bibinfo {pages} {443} (\bibinfo {year} {1992})}\BibitemShut
  {NoStop}%
\bibitem [{\citenamefont {Dennis}\ \emph {et~al.}(2002)\citenamefont {Dennis},
  \citenamefont {Kitaev}, \citenamefont {Landahl},\ and\ \citenamefont
  {Preskill}}]{Dennis2001}%
  \BibitemOpen
  \bibfield  {author} {\bibinfo {author} {\bibfnamefont {E.}~\bibnamefont
  {Dennis}}, \bibinfo {author} {\bibfnamefont {A.}~\bibnamefont {Kitaev}},
  \bibinfo {author} {\bibfnamefont {A.}~\bibnamefont {Landahl}}, \ and\
  \bibinfo {author} {\bibfnamefont {J.}~\bibnamefont {Preskill}},\ }\href
  {\doibase 10.1063/1.1499754} {\bibfield  {journal} {\bibinfo  {journal} {J.
  Math. Phys.}\ }\textbf {\bibinfo {volume} {43}},\ \bibinfo {pages} {4452}
  (\bibinfo {year} {2002})},\ \Eprint {http://arxiv.org/abs/quant-ph/0110143v1}
  {quant-ph/0110143v1} \BibitemShut {NoStop}%
\bibitem [{\citenamefont {Milsted}\ and\ \citenamefont
  {Osborne}(2016)}]{Milsted2016}%
  \BibitemOpen
  \bibfield  {author} {\bibinfo {author} {\bibfnamefont {A.}~\bibnamefont
  {Milsted}}\ and\ \bibinfo {author} {\bibfnamefont {T.~J.}\ \bibnamefont
  {Osborne}},\ }\href@noop {} {\  (\bibinfo {year} {2016})},\ \Eprint
  {http://arxiv.org/abs/1604.01979v1} {1604.01979v1} \BibitemShut {NoStop}%
\bibitem [{\citenamefont {Witten}(1989)}]{Witten1989}%
  \BibitemOpen
  \bibfield  {author} {\bibinfo {author} {\bibfnamefont {E.}~\bibnamefont
  {Witten}},\ }\href {http://projecteuclid.org/euclid.cmp/1104178138}
  {\bibfield  {journal} {\bibinfo  {journal} {Comm. Math. Phys.}\ }\textbf
  {\bibinfo {volume} {121}},\ \bibinfo {pages} {351} (\bibinfo {year}
  {1989})}\BibitemShut {NoStop}%
\bibitem [{\citenamefont {Kitaev}(2003)}]{Kitaev2003}%
  \BibitemOpen
  \bibfield  {author} {\bibinfo {author} {\bibfnamefont {A.}~\bibnamefont
  {Kitaev}},\ }\href {\doibase http://dx.doi.org/10.1016/S0003-4916(02)00018-0}
  {\bibfield  {journal} {\bibinfo  {journal} {Ann. Phys.}\ }\textbf {\bibinfo
  {volume} {303}},\ \bibinfo {pages} {2 } (\bibinfo {year} {2003})}\BibitemShut
  {NoStop}%
\bibitem [{\citenamefont {Levin}\ and\ \citenamefont {Wen}(2005)}]{Levin2005}%
  \BibitemOpen
  \bibfield  {author} {\bibinfo {author} {\bibfnamefont {M.~A.}\ \bibnamefont
  {Levin}}\ and\ \bibinfo {author} {\bibfnamefont {X.-G.}\ \bibnamefont
  {Wen}},\ }\href {\doibase 10.1103/PhysRevB.71.045110} {\bibfield  {journal}
  {\bibinfo  {journal} {Phys. Rev. B}\ }\textbf {\bibinfo {volume} {71}},\
  \bibinfo {pages} {045110} (\bibinfo {year} {2005})}\BibitemShut {NoStop}%
\bibitem [{Note3()}]{Note3}%
  \BibitemOpen
  \bibinfo {note} {The circuit that creates the \protect \emph {entire state},
  as opposed to a single row, is not finite-depth; it scales linearly with the
  system size. This is unavoidable due to the existence of topological quantum
  order\cite {Bravyi2006,Koenig2014}.}\BibitemShut {Stop}%
\bibitem [{\citenamefont {Kim}(2017)}]{Kim2017}%
  \BibitemOpen
  \bibfield  {author} {\bibinfo {author} {\bibfnamefont {I.~H.}\ \bibnamefont
  {Kim}},\ }\href@noop {} {\bibfield  {journal} {\bibinfo  {journal} {in
  preparation}\ } (\bibinfo {year} {2017})}\BibitemShut {NoStop}%
\bibitem [{\citenamefont {Evenbly}\ and\ \citenamefont
  {Vidal}(2009)}]{Evenbly2009}%
  \BibitemOpen
  \bibfield  {author} {\bibinfo {author} {\bibfnamefont {G.}~\bibnamefont
  {Evenbly}}\ and\ \bibinfo {author} {\bibfnamefont {G.}~\bibnamefont
  {Vidal}},\ }\href {\doibase 10.1103/PhysRevB.79.144108} {\bibfield  {journal}
  {\bibinfo  {journal} {Phys. Rev. B}\ }\textbf {\bibinfo {volume} {79}},\
  \bibinfo {pages} {144108} (\bibinfo {year} {2009})}\BibitemShut {NoStop}%
\bibitem [{\citenamefont {Verstraete}\ \emph {et~al.}(2008)\citenamefont
  {Verstraete}, \citenamefont {Cirac},\ and\ \citenamefont
  {Murg}}]{Verstraete2009}%
  \BibitemOpen
  \bibfield  {author} {\bibinfo {author} {\bibfnamefont {F.}~\bibnamefont
  {Verstraete}}, \bibinfo {author} {\bibfnamefont {J.~I.}\ \bibnamefont
  {Cirac}}, \ and\ \bibinfo {author} {\bibfnamefont {V.}~\bibnamefont {Murg}},\
  }\href {\doibase 10.1080/14789940801912366} {\bibfield  {journal} {\bibinfo
  {journal} {Adv. Phys.}\ }\textbf {\bibinfo {volume} {57}},\ \bibinfo {pages}
  {143} (\bibinfo {year} {2008})},\ \Eprint {http://arxiv.org/abs/0907.2796v1}
  {0907.2796v1} \BibitemShut {NoStop}%
\bibitem [{\citenamefont {Devoret}\ and\ \citenamefont
  {Schoelkopf}(2013)}]{Devoret2013}%
  \BibitemOpen
  \bibfield  {author} {\bibinfo {author} {\bibfnamefont {M.~H.}\ \bibnamefont
  {Devoret}}\ and\ \bibinfo {author} {\bibfnamefont {R.~J.}\ \bibnamefont
  {Schoelkopf}},\ }\href@noop {} {\bibfield  {journal} {\bibinfo  {journal}
  {Science}\ }\textbf {\bibinfo {volume} {339}},\ \bibinfo {pages} {1169}
  (\bibinfo {year} {2013})}\BibitemShut {NoStop}%
\bibitem [{\citenamefont {Cubitt}\ \emph {et~al.}(2015)\citenamefont {Cubitt},
  \citenamefont {Lucia}, \citenamefont {Michalakis},\ and\ \citenamefont
  {Perez-Garcia}}]{Cubitt2013}%
  \BibitemOpen
  \bibfield  {author} {\bibinfo {author} {\bibfnamefont {T.~S.}\ \bibnamefont
  {Cubitt}}, \bibinfo {author} {\bibfnamefont {A.}~\bibnamefont {Lucia}},
  \bibinfo {author} {\bibfnamefont {S.}~\bibnamefont {Michalakis}}, \ and\
  \bibinfo {author} {\bibfnamefont {D.}~\bibnamefont {Perez-Garcia}},\ }\href
  {\doibase 10.1007/s00220-015-2355-3} {\bibfield  {journal} {\bibinfo
  {journal} {Comm. Math. Phys.}\ }\textbf {\bibinfo {volume} {337}},\ \bibinfo
  {pages} {1275} (\bibinfo {year} {2015})},\ \Eprint
  {http://arxiv.org/abs/1303.4744v4} {1303.4744v4} \BibitemShut {NoStop}%
\bibitem [{\citenamefont {Lucia}\ \emph {et~al.}(2015)\citenamefont {Lucia},
  \citenamefont {Cubitt}, \citenamefont {Michalakis},\ and\ \citenamefont
  {P\'erez-Garc\`ia}}]{Lucia2014}%
  \BibitemOpen
  \bibfield  {author} {\bibinfo {author} {\bibfnamefont {A.}~\bibnamefont
  {Lucia}}, \bibinfo {author} {\bibfnamefont {T.~S.}\ \bibnamefont {Cubitt}},
  \bibinfo {author} {\bibfnamefont {S.}~\bibnamefont {Michalakis}}, \ and\
  \bibinfo {author} {\bibfnamefont {D.}~\bibnamefont {P\'erez-Garc\`ia}},\
  }\href {\doibase 10.1103/PhysRevA.91.040302} {\bibfield  {journal} {\bibinfo
  {journal} {Phys. Rev. A}\ }\textbf {\bibinfo {volume} {91}},\ \bibinfo
  {pages} {040302} (\bibinfo {year} {2015})},\ \Eprint
  {http://arxiv.org/abs/1409.7809v3} {1409.7809v3} \BibitemShut {NoStop}%
\bibitem [{\citenamefont {Bravyi}\ \emph {et~al.}(2006)\citenamefont {Bravyi},
  \citenamefont {Hastings},\ and\ \citenamefont {Verstraete}}]{Bravyi2006}%
  \BibitemOpen
  \bibfield  {author} {\bibinfo {author} {\bibfnamefont {S.}~\bibnamefont
  {Bravyi}}, \bibinfo {author} {\bibfnamefont {M.~B.}\ \bibnamefont
  {Hastings}}, \ and\ \bibinfo {author} {\bibfnamefont {F.}~\bibnamefont
  {Verstraete}},\ }\href {\doibase 10.1103/PhysRevLett.97.050401} {\bibfield
  {journal} {\bibinfo  {journal} {Phys. Rev. Lett.}\ }\textbf {\bibinfo
  {volume} {97}},\ \bibinfo {pages} {050401} (\bibinfo {year}
  {2006})}\BibitemShut {NoStop}%
\bibitem [{\citenamefont {K\"onig}\ and\ \citenamefont
  {Pastawski}(2014)}]{Koenig2014}%
  \BibitemOpen
  \bibfield  {author} {\bibinfo {author} {\bibfnamefont {R.}~\bibnamefont
  {K\"onig}}\ and\ \bibinfo {author} {\bibfnamefont {F.}~\bibnamefont
  {Pastawski}},\ }\href {\doibase 10.1103/PhysRevB.90.045101} {\bibfield
  {journal} {\bibinfo  {journal} {Phys. Rev. B}\ }\textbf {\bibinfo {volume}
  {90}},\ \bibinfo {pages} {045101} (\bibinfo {year} {2014})}\BibitemShut
  {NoStop}%
\bibitem [{\citenamefont {Stinespring}(1955)}]{Stinespring1955}%
  \BibitemOpen
  \bibfield  {author} {\bibinfo {author} {\bibfnamefont {W.~F.}\ \bibnamefont
  {Stinespring}},\ }\href {http://www.jstor.org/stable/2032342} {\bibfield
  {journal} {\bibinfo  {journal} {Proc. Amer. Math. Soc.}\ }\textbf {\bibinfo
  {volume} {6}},\ \bibinfo {pages} {211} (\bibinfo {year} {1955})}\BibitemShut
  {NoStop}%
\bibitem [{\citenamefont {Lieb}(1973)}]{Lieb1973}%
  \BibitemOpen
  \bibfield  {author} {\bibinfo {author} {\bibfnamefont {E.~H.}\ \bibnamefont
  {Lieb}},\ }\href {\doibase 10.1063/1.1666274} {\bibfield  {journal} {\bibinfo
   {journal} {J. Math. Phys.}\ }\textbf {\bibinfo {volume} {14}},\ \bibinfo
  {pages} {1938} (\bibinfo {year} {1973})}\BibitemShut {NoStop}%
\bibitem [{\citenamefont {Fawzi}\ and\ \citenamefont
  {Renner}(2015)}]{Fawzi2015}%
  \BibitemOpen
  \bibfield  {author} {\bibinfo {author} {\bibfnamefont {O.}~\bibnamefont
  {Fawzi}}\ and\ \bibinfo {author} {\bibfnamefont {R.}~\bibnamefont {Renner}},\
  }\href {\doibase 10.1007/s00220-015-2466-x} {\bibfield  {journal} {\bibinfo
  {journal} {Comm. Math. Phys.}\ }\textbf {\bibinfo {volume} {340}},\ \bibinfo
  {pages} {575} (\bibinfo {year} {2015})},\ \Eprint
  {http://arxiv.org/abs/1410.0664} {arXiv:1410.0664} \BibitemShut {NoStop}%
\end{thebibliography}%
\appendix

\section{1D simulator can represent a gapped 2D state\label{section:faithfulness}}
The states obeying Eq.\ref{eq:TEE} have a form of Eq.\ref{eq:fcs}, where the underlying quantum channels are local. That is, the channels at each $t$ can be decomposed into finite-depth quantum channels which are localized in space. Because every channel can be recast into a unitary operation followed by partial trace\cite{Stinespring1955}, such states can be prepared by applying a  finite-depth quantum circuit and taking a partial trace over some of the qubits.

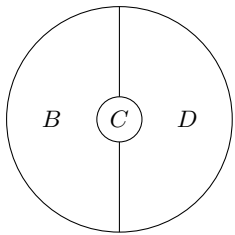
\begin{figure}[h]
	\begin{tikzpicture}
		\draw (0,0) circle (0.3);
		\draw (0,0) circle (1.5);
		\draw (0,0.3)--(0,1.5);
		\draw (0,-0.3)--(0,-1.5);
		\draw (0,0) node {$C$};
		\draw (-0.9,-0) node {$B$};
		\draw (0.9,0) node {$D$};
	\end{tikzpicture}
	\caption{The choice of $B,C$ and $D$ for Eq.\ref{eq:local_constraint}.\label{fig:local_constraint}}
\end{figure}

The area law implies an identity that is universal to \emph{all} topological quantum field theories:
\begin{equation}
	S(BC) - S(B) + S(CD) - S(D) \approx 0, \label{eq:local_constraint}
\end{equation}
where the subsystems are depicted in FIG.\ref{fig:local_constraint}\cite{Kim2014}. 
 In Ref.\cite{Kim2014}, it was shown that Eq.\ref{eq:local_constraint} implies the following:
\begin{equation}
	S(AB) + S(BC) - S(B) - S(ABC)\approx 0,
\end{equation}
where the subsystems are depicted in FIG.\ref{fig:cmi}. This follows from the so called weak monotonicity, also known to be equivalent to the strong subadditivity of entropy\cite{Lieb1973}. The technical statement is that
\begin{equation}
  S(AB)- (B) + S(AC) - S(C) \geq 0 \label{eq:wm}
\end{equation}
for any tripartite quantum state over $A,B,$ and $C$. Note that 
\begin{equation}
  \label{eq:temp}
  \begin{aligned}
    S(AB) + S(BC) - S(B) - S(ABC) \\ \leq S(BC) - S(B) + S(CD) - S(D),
    \end{aligned}
\end{equation}
by choosing the subsystem $A$ in Eq.\ref{eq:wm} to be the subsystem $C$ in Eq.\ref{eq:temp}, $B$ in Eq.\ref{eq:wm} to be $AB$ in Eq.\ref{eq:temp}, and $C$ in Eq.\ref{eq:wm} to be $D$.

\begin{figure}[h]
	\begin{tikzpicture}
		\draw (-3,0)--(3,0)--(3,1)--(-3,1)--(-3,0);
		\draw (-0.25,1)--(0.25,1)--(0.25,1.5)--(-0.25,1.5)--(-0.25,1);
		\draw (-0.75,1)--(-0.75,0.5)--(0.75,0.5)--(0.75,1);
		\draw (0,1.25) node {$C$};
		\draw (0,0.75) node {$B$};
		\draw (0,0.25) node {$A$};
	\end{tikzpicture}
	\caption{A state on a slab can be extended to a state that contains one more qubit by applying a local channel, supported on $BC$.\label{fig:cmi}}
\end{figure}
The theorem of Fawzi and Renner\cite{Fawzi2015} then implies that there exists a quantum channel $\Phi$ acting on $BC$ such that 
\begin{equation}
	\rho^{ABC} \approx \Phi (\rho^{AB}).
\end{equation}
By Stinespring's theorem\cite{Nielsen2011}, the action of such channel can be reproduced by introducing an auxiliary degrees of freedom of bounded dimension, applying a unitary on the enlarged system, and tracing them out. These auxiliary degrees of freedom are nothing but the sink qubits of our proposal.

The argument outlined above applies in the bulk of the 2D system, and a completely analogous argument works for the physical boundary as well, provided that the boundary is gapped. In this case, we can posit that the entanglement entropy of a region that includes the boundary has the following form:
\begin{equation}
  S(A) = \alpha l -\gamma' + \cdots,
\end{equation}
where $\gamma'$ is a term that only depends on the topology of $A$. Then,
\begin{equation}
	S(BC) - S(B) + S(CD) - S(D) \approx 0, \label{eq:local_constraint2}
\end{equation}
for the following choice of subsystems.
\begin{figure}[h]
	\begin{tikzpicture}
          \draw [domain=180:360] plot ({1.5*cos(\x)}, {1.5*sin(\x)});
          \draw [domain=180:360] plot ({0.4*cos(\x)}, {0.4*sin(\x)});
          \draw (0,-0.4)--(0,-1.5);
          \draw (-1.5,0)--(1.5,0);
          \draw (0,0) node[below] {$C$};
          \draw (-0.9,-0) node[below] {$B$};
          \draw (0.9,0) node[below] {$D$};
	\end{tikzpicture}
	\caption{The choice of $A,B,$ and $C$ for Eq.\ref{eq:local_constraint2}. The flat part represents the physical boundary of the 2D system.\label{fig:local_constraint2}}
\end{figure}
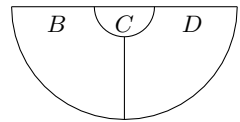
Again following Ref.\cite{Kim2014}, Eq.\ref{eq:local_constraint2} implies the following:
\begin{equation}
	S(AB) + S(BC) - S(B) - S(ABC)\approx 0,
\end{equation}
where the subsystems are depicted in FIG.\ref{fig:cmi2}
\begin{figure}[h]
	\begin{tikzpicture}
		\draw (-3,0)--(3,0)--(3,1)--(-3,1)--(-3,0);
		\draw (-3,1)--(-2.5,1)--(-2.5,1.5)--(-3,1.5)--(-3,1);
		\draw (-3,1)--(-3,0.5)--(-1.5,0.5)--(-1.5,1);
		\draw (-2.75,1.25) node {$C$};
		\draw (-2.25,0.75) node {$B$};
		\draw (0,0.25) node {$A$};
	\end{tikzpicture}
	\caption{A state on a slab can be extended to a state that contains one more qubit by applying a local channel, supported on $BC$.\label{fig:cmi2}}
\end{figure}
The theorem of Fawzi and Renner\cite{Fawzi2015} then implies that there exists a quantum channel $\Phi$ acting on $BC$ such that 
\begin{equation}
	\rho^{ABC} \approx \Phi (\rho^{AB}).
\end{equation}

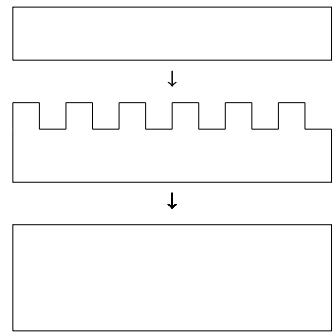
\begin{figure}[h]
	\resizebox{0.5\columnwidth}{!}{
	\begin{tikzpicture}
		\draw (-3,0)--(3,0)--(3,1)--(-3,1)--(-3,0);
		\draw[thick,->] (0,-0.2)--(0,-0.5);
		\begin{scope}[yshift=-2.3cm]	
		\draw (-3,1)--(-3,0)--(3,0)--(3,1);
		\foreach \x in {-3,-2.5,...,2.5} {
			\draw (\x,1)--(\x,1.5);
		}
		\foreach \x in {-3,-2,...,2}{
			\draw (\x,1.5)--(\x+0.5,1.5);
		}
		\foreach \x in {-2.5,-1.5,...,2.5}{
			\draw (\x,1)--(\x+0.5,1);
		\draw[thick,->] (0,-0.2)--(0,-0.5);
		}
		\end{scope}
		\begin{scope}[yshift=-5.1cm]
			\draw (-3,0)--(3,0)--(3,2)--(-3,2)--(-3,0);
		\end{scope}
	\end{tikzpicture}}
	\caption{A state on a slab can be extended to a state on a larger slab by a low-depth quantum channel.\label{fig:channel_decomposition}}
\end{figure}

One can recursively use this logic to show that a sequence of local quantum channels can extend a state on a slab to another state on a slab that is larger by a unit width. Because each of these channels are localized in space, they can be rearranged in such a way that the entire process can be compressed into finite number of local quantum channels that run in parallel; see FIG.\ref{fig:channel_decomposition}

Now, consider the following physical procedure. We first create and store the state of the first slab in the bath-sink composite in such a way that some of the sink qubits store the first row of the slab. We swap these qubits with the first row of the system qubits. At this point, the sink qubits that stored the first row is in some fixed state. Swap this fixed state with the qubits that store the second row, and then with the third row etc. At the end of this procedure, the sink qubits that stored the first row now stores the second row, and we are left with the fixed state. The bath-sink composite now stores a slab of width $w-1$ and a fixed state upon which we can extend the state. Apply the procedure explained in FIG.\ref{fig:channel_decomposition}, extending the state on a slab of width $w-1$ to a state on a slab of width $w$, ranging from the second row to the $(w+1)$th row. This completes the argument.

\end{document}